\documentstyle[aps,prl]{revtex}
\begin{document}
\draft
\author{Robert de Mello Koch and Jo\~ao P. Rodrigues,
}
\address{Department of Physics and Centre for Non Linear Studies,\\ 
University of
the Witwatersrand, Wits 2050, South Africa}
\title{\hspace{2.5in} CNLS-97-01\\
Classical integrability of chiral $QCD_{2}$ and classical curves.}

%\twocolumn[
\maketitle
\begin{minipage}{\textwidth}
\begin{quotation}
\begin{abstract}
In this letter, classical chiral $QCD_{2}$ is studied in the lightcone
gauge $A_{-}=0$. The once integrated equation of motion for the 
current
is shown to be of the Lax form, which demonstrates an infinite number
of conserved quantities. Specializing to gauge group $SU(2)$, we show
that solutions to the classical equations of motion can be identified
with a very large class of curves. We demonstrate this correspondence
explicitly for two solutions. The classical fermionic fields 
associated
with these currents are then obtained. 
\bigskip
\end{abstract}
\end{quotation}
\end{minipage}

1) In the early seventies, a number of 1+1 dimensional field theories 
were shown to be integrable~\cite{thacker}. It was soon realized that 
the S matrices of these theories could be computed exactly~\cite{Zam} 
as well as the eigenfunctions and spectrum of the Hamiltonian 
~\cite{And}. 
There have been recent claims using 
non-abelian bosonization that two dimensional $QCD$ with massless 
quarks in the fundamental representation is integrable 
~\cite{Abdalla}.

There are a certain number of properties usually associated with  
integrable two dimensional field theories: their classical equations 
of motion can be recast as a Lax form, from which the existence of an 
infinite number of conservation laws can be established, and if these 
survive quantization they in turn ultimately imply an elastic S-
matrix 
with no particle production. These theories are also 
usually associated with the existence of a large number of soliton-
like solutions of the classical equations of motion.

The specific way in which the form of $QCD_{2}$ integrability might 
be 
reconciled with the considerable existing evidence of particle 
production~\cite{Coote},~\cite{Stau} in the large $N_{c}$ expansion 
of 
$QCD_{2}$~\cite{tHooft} remains to be explained. 
\footnote{There are 1+1 string theories 
with infinite dimensional symmetries 
and particle production~\cite{Pol}, but this results from 
momentum non-conservation.}$^{,}$\footnote{For a recent attempt to
reconcile these distinct aspects of $QCD_{2}$, see 
reference~\cite{abdalla}.}

In this letter, we investigate the existence of exact solutions to 
the 
classical equations of motion of chiral $QCD_{2}$ in the gauge $A_{-
}=0$ 
and in light cone coordinates. We start by obtaining the classical 
equations of motion for colored $SU(N)$ currents and recast them in a 
Lax 
form . Specializing to $SU(2)$, we show 
that solutions to these equations of 
motion can be identified with 
a very large class of curves described by Frenet-Serret 
equations. 
They are also invariant under a continuos symmetry.
We briefly describe two particular solutions, one of which can be 
made to
acquire the quantum numbers of a baryon.

We then proceed to obtain the classical fermion fields associated 
with these currents. Although the relevance of classical solutions 
for fermionic theories might not be obvious, they have been used in
the past by Papanicolaou ~\cite{Pap}, to obtain equal time large $N$
bilocals. We implement a similar
construction using color for 'tHooft's solution. 

2) We work in lightcone co-ordinates \footnote{ Our conventions are 
$x^{\pm}={1\over\sqrt{2}}(x^{0}\pm x^{1})=x_{\mp}$ and 
$\gamma^{0}=\sigma_{1}$,
$\gamma^{1}=i\sigma_{2}$} and in the axial gauge $A_{-}=0$. For 
quarks $\psi_{a}=[\psi_{-a}, \psi_{+a}]^{T}$ in the
fundamental representation of $SU(N)$
($a=1,2,...N$), the Lagrangian density is given by

\begin{eqnarray}
\nonumber
{\cal L}&&= {1\over 2}Tr(\partial_{-}A_{+})^{2}+
i\sqrt{2}{\psi_{+a}}^{\dagger}\partial_{-}\psi_{+a}+ 
\sqrt{2}{\psi_{-a}}^{\dagger}(i\delta_{ab}\partial_{+}
+gA_{+ab})\psi_{-b}\\
&&-m(\psi_{-a}^{\dagger}\psi_{+a}+\psi^{\dagger}_{+a}\psi_{-a}).
\label{eqnofmotion}
\end{eqnarray}

With $x^{+}$ as "time", the fields $A_{+}$ and $\psi_{+}$ are not 
dynamical. Solving their constraint equations, we let
$\psi_{-a}\to\psi_{a}$, $m\to 0$ and obtain ($x=(x^{+},x^{-})$):

\begin{eqnarray}
\nonumber
{\cal L}&&= 
i\sqrt{2}{\psi_{a}}^{\dagger}(x)\partial_{+}\psi_{a}(x)+
{g^{2}\over 2} \int dy^{-}\times \\
\nonumber
&&\times{\psi_{a}}^{\dagger}(x)\psi_{b}(x) |x^{-}-y^{-}|
{\psi_{b}}^{\dagger}(x^{+},y^{-})\psi_{a}(x^{+},y^{-})
-{g^{2}\over 2N} \int dy^{-}\times \\
&&\times{\psi_{a}}^{\dagger}(x)\psi_{a}(x) |x^{-}-y^{-}|
{\psi_{b}}^{\dagger}(x^{+},y^{-})\psi_{b}(x^{+},y^{-}).
\label{QCDLagrangian}
\end{eqnarray}

It is known that the $m=0$ limit is singular in two dimensions.
This is, however, the limit considered in [4] and in any case
we will restrict ourselves in this paper mainly to classical
considerations.
The classical equation of motion can be written as

\begin{equation}
i\partial_{+}\psi_{a}(x)={g^{2}\over\sqrt{2}}\sigma_{ab}(x)
\psi_{b}(x)
\label{EqnMot}
\end{equation}

where the potential $\sigma_{ab}$ is given by

\begin{equation}
\sigma_{ab}(x)= \int dy^{-}|x^{-}-y^{-}|\big[
\psi_{a}(x^{+},y^{-})
\psi_{b}^{\dagger}(x^{+},y^{-})-{1\over N} \delta_{ab}\psi_{c}
(x^{+},y^{-})\psi^{\dagger}_{c}(x^{+},y^{-})\big].
\label{Sigm}
\end{equation}

The currents are related to the potential by

\begin{equation}
\partial_{-}^{2}\sigma_{ab}(x)=2\Big(\psi_{a}(x)\psi^{\dagger}_{b}
(x) -{1\over N}\delta_{ab}\psi_{c}(x)\psi^{\dagger}_{c}(x)\Big),
\label{SigmPsi}
\end{equation}

and it follows from ~(\ref{EqnMot}) that they satisfy

\begin{equation}
\partial_{+}\partial_{-}^{2}\sigma ={ig^{2}\over\sqrt{2}}
[\partial_{-}^{2}\sigma ,\sigma].
\label{FLaxExplicit}
\end{equation}

Equation~(\ref{FLaxExplicit}) is after all the classical covariant 
conservation 
equation for the $SU(N)$ currents with an "external" gauge potential 
expressed via Gauss' Law in terms of the fermionic currents 
themselves. 
From this point of view, the presence of the commutator on the right 
hand 
side of~(\ref{FLaxExplicit}) is entirely natural. However, 
interpreting 
$x^{+}$ 
as time, equation~(\ref{FLaxExplicit}) takes the form of an (ultra-
local) 
Lax pair. It is also a total $x^{-}$ derivative, an observation that 
forms the basis of the analysis of~\cite{Abdalla}. \footnote{ The 
equation 
considered in~\cite{Abdalla} has an extra term associated with the 
chiral 
anomaly, resulting from integrating out the fermionic fields as part 
of the sequence of change of variables used in the approach.}

To obtain the time dependence of $\partial_{-}\sigma_{ab}$, we write 
(using (4) - (6)) 

\begin{eqnarray}
\nonumber
i\partial_{+}\partial_{-}\sigma_{ab}(x)
&&=i\partial_{+}\int dy^{-} \epsilon (x^{-}-y^{-})\big[
\psi_{a}(x^{+},y^{-})\psi_{b}^{\dagger}(x^{+},y^{-})
-{1\over N}\delta_{ab}
\psi_{c}(x^{+},y^{-})\psi_{c}^{\dagger}(x^{+},y^{-})\big]\\
&&={ig^{2}\over 2\sqrt{2}}
\int dy^{-}\epsilon (x^{-}-y^{-})[\sigma,\partial_{-
}^{2}\sigma]_{ab}.
\label{DerivSigm}
\end{eqnarray}

Noticing that 
${\partial_{-}[\partial_{-}\sigma(x),\sigma(x)]
=[\partial^{2}_{-}\sigma(x),\sigma(x)]}$, 
one can integrate by parts, and provided the boundary term vanishes, 
one obtains

\begin{equation}
{i\partial_{+}\partial_{-}\sigma_{ab}(x)={g^{2}\over\sqrt{2}}
[\sigma(x),\partial_{-}\sigma(x)]_{ab}}.
\label{DerivSigmTwo}
\end{equation}

This equation is the integrated version of equation (6), and the 
above argument explains the precise conditions under which it
indeed follows from equation (6).

Both equations (6) and (8) are of the Lax form and highly unusual
for a two dimensional system. They imply that  

\begin{equation}
{\partial_{+}Tr(\partial_{-}\sigma (x))^{n}=0; 
\quad \partial_{+}Tr(\partial_{-}^{2}\sigma(x))^{n}=0}
\label{ChargesTwo}
\end{equation}

for every "space" point $x^{-}$, generating an infinite number of 
conserved charges. One should notice that $\partial_{-}\sigma$
depends nonlocally on the quark fields. Nonlocal conservation laws
have been discussed in~\cite{AntPap}. Due to the fact that one is
dealing with a gauge theory, one might be concerned that some of
the properties described above are artifacts of the gauge choice. An
example has been discussed in [18]. However, certainly
the infinite set of local conserved quantities in (9) are gauge 
invariant.

3) We observe that the equations of motion (6) and (8) are invariant 
under
the continuos symmetry $x^{-}\to f(x^{-}).$ This is unexpected
since the action does not have such a symmetry (we will describe 
later how 
this symmetry is implemented at the level of the classical fermion 
fields)
\footnote{As an example, it is easy to see from (4) that under 
$\psi\to\psi'(x)={1\over a}\psi(x^{+},ax^{-})$,
$\sigma'(x)=\sigma(x^{+},ax^{-})$ and that this transformation can 
not be a
symmetry of the action since it does not respect the conformal (scale)
weight of the fermion field.}. This situation where the equations of 
motion 
have more symmetry than the action from which they are derived is 
somewhat
unique.

From now on, we specialize to $SU(2)$ and write

\begin{equation}
\sigma_{ab}(x)={1\over 2}f^{i}(x)\sigma^{i}_{ab}
\label{ExpSigm}
\end{equation}

where the $\sigma^{i}$ are the standard Pauli matrices. Equation (8)
becomes

\begin{equation}
{\partial_{+}\partial_{-}f^{i}={g^{2}\over\sqrt{2}}
\epsilon_{ijk}\partial_{-}f^{j}f^{k}}.
\label{fdyn}
\end{equation}

Since from (9) $\partial_{-}f^{i}\partial_{-}f^{i}$ is "time" 
independent,
we use the $x^{-}$ diffeomorphism invariance of (11) to move to an
arclength co-ordinate \footnote{This is in the manifold of the Lie 
algebra
of course.} $s$ for which $\partial_{s}f^{i}\partial_{s}f^{i}=1.$

We note the similarity between the equations of motion

\begin{equation}
{\partial_{+}\partial_{s}f^{i}={g^{2}\over\sqrt{2}}
\epsilon_{ijk}\partial_{s}f^{j}f^{k},\quad \partial_{s}f^{i}
\partial_{s}f^{i}=1}
\label{fdynArc}
\end{equation}

and those of the continuos Heisenberg spin chain~\cite{AntPap}

\begin{equation}
{\partial_{t}S^{i}=
\epsilon_{ijk}S^{j}\partial_{x}^{2}S^{k},\quad S^{i}S^{i}=s(s+1).}
\label{HSC}
\end{equation}

The normal, binormal and osculating normal respectively

\begin{equation}
{e_{1}^{i}=\partial_{s}f^{i}, \quad
e_{2}^{i}={1\over\kappa}\partial_{s}^{2}f^{i}}, \quad
e_{3}^{i}={1\over\kappa}\epsilon_{ijk}
\partial_{s}f^{j}\partial_{s}^{2}f^{k},
\label{Nor}
\end{equation}

(with the curvature given by
$\kappa^{2} =\partial_{s}^{2}f^{i}\partial_{s}^{2}f^{i}$) satisfy,
using (12),

\begin{eqnarray}
\nonumber
\partial_{+}e_{1}^{i}=
{g^{2}\over\sqrt{2}}\epsilon_{ijk}f^{j}e_{1}^{k}&&,
\qquad
\partial_{+}e_{2}^{i}=-{1\over\kappa}(\partial_{+}\kappa)e_{2}^{i}+
{g^{2}\over\sqrt{2}}\epsilon_{ijk}f^{j}e_{2}^{k}\\
\partial_{+}e_{3}^{i}&&=-{1\over\kappa}(\partial_{+}\kappa)e_{3}^{i}+
{g^{2}\over\sqrt{2}}\epsilon_{ijk}f^{j}e_{3}^{k}.
\label{NewEqns}
\end{eqnarray}

They are also known to satisfy the Frenet-Serret equations

\begin{equation}
\partial_{s}e_{1}^{i}=\kappa e_{2}^{i}\quad
\partial_{s}e_{2}^{i}=\tau e_{3}^{i}-\kappa e_{1}^{i}\quad
\partial_{s}e_{3}^{i}=-\tau e_{2}^{i}.
\label{SF}
\end{equation}

with the torsion $\tau$ defined as 
$\kappa^{2}\tau=\epsilon_{ijk}f^{i}_{,s}f^{j}_{,ss}f^{k}_{,sss}.$ 
Requiring
(15) and (16) to be compatible we immediatly obtain

\begin{equation}
\partial_{+}\kappa =0, \quad \partial_{+}\tau=\sqrt{2}g^{2}, \quad
\partial_{+}(e_{n})^{i}={g^{2}\over\sqrt{2}}
\epsilon_{ijk}f^{j}e_{n}^{k},
\label{compatability}
\end{equation}

i.e. $\kappa =f(x^{-}),$ $\tau={g^{2}\over\sqrt{2}}x^{+}+h(x^{-}),$
with $f,g$ arbitrary. Therefore, all curves with a torsion
rising linearly in "time" $x^{+}$ (corresponding to a constant 
contribution
to the torsion at an equal "time" slice) are solutions of (12). 
Conversely,
this very large class of curves is equivalently described by equation 
(12).
It is known that for a given torsion and curvature a classical curve 
is
specified by a solution to the Ricatti equation~\cite{Curve}. Its 
equivalent
description via equation (12) indicates that it may be possible in 
some 
cases to use inverse scattering methods on equation (12) to obtain
solutions of the Ricatti equation.

The connection between solitons and curves has been discussed by a 
number
of authors~\cite{Has}. In particular, Lamb~\cite{Lamb} in connection 
with the
Sine-Gordon equation, has described a curve of curvature 
$\kappa =2asech (as)$ and torsion $\tau =$ constant. This curve upon
identifying $\tau={g^{2}\over\sqrt{2}}x^{+}$ provides a soliton 
solution
of (3.2). Explicitly

\begin{eqnarray}
f^{1}&&={-2a \over {g^{4}\over 2}x^{+2}+a^{2}}sech(as) 
sin({g^{2}\over \sqrt{2}}x^{+}s)
\cr\noalign{\vskip 0.2truecm}
f^{2}&&= {2a \over {g^{4}\over 2}x^{+2}+a^{2}}sech(as) 
cos({g^{2}\over \sqrt{2}}x^{+}s)
\cr\noalign{\vskip 0.2truecm}
f^{3}&&= s-{2a\over {g^{4}\over 2}x^{+2}+a^{2}}tanh(as).
\label{CurveSoln}
\end{eqnarray}

In a separate communication, we will show that this is the 
reflectionless
single bound state eigenvalue solution to the Gelfand-Levitan-
Marchenko
equation, arising in a systematic inverse scattering treatment of 
chiral
$QCD_{2}$.

We also state the solutions with constant curvature $\kappa$ and 
$\tau={g^{2}\over\sqrt{2}}x^{+}:$

\begin{eqnarray}
f^{1}&&={\kappa \over {g^{4}x_{+}^{2}\over 2}+\kappa^{2}} 
sin(\big[\sqrt{\kappa^{2}+{g^{4}x^{+2}\over 2}}\big]s)
\cr\noalign{\vskip 0.2truecm}
f^{2}&&={\kappa \over {g^{4}x_{+}^{2}\over 2}+\kappa^{2}} 
cos(\big[\sqrt{\kappa^{2}+{g^{4}x^{+2}\over 2}}\big]s)
\cr\noalign{\vskip 0.2truecm}
f^{3}&&=-{g^{2}x^{+}s\over \sqrt{2}\sqrt{\kappa^{2}+{g^{4}x^{+2}\over 
2}} }.
\label{SecondCurveSoln}
\end{eqnarray}

4) Provided the fermion fields are treated 
classically, as we will do in the
following, it follows from equation (5) for 
the $SU(2)$ case that

\begin{equation}
\partial_{-}^{2}\sigma_{ab}\psi_{b}=
(\psi_{c}\psi_{c}^{\dagger})\psi_{a}
\equiv\lambda\psi_{a}.
\label{EigenvalueEqn}
\end{equation}

The fermion number density $\lambda$ is easily 
seen to be related to the curvature
$\kappa$ by $\kappa=2\lambda$, and is 
therefore $x^{+}$ independent.
Equation ~(\ref{EigenvalueEqn}) determines $\psi_{a}$ 
up to a phase, which we now 
show can always be chosen so that ~(\ref{EqnMot}) 
is satisfied: differentiating
equation (20) with respect to $x^{+}$ and using 
equation (6), we readily
obtain (in matrix notation)

\begin{equation}
{\partial_{-}^{2}\sigma(i\partial_{+}-{g^{2}\over\sqrt{2}}
\sigma)\psi=\lambda(i\partial_{+}-{g^{2}\over\sqrt{2}}\sigma)\psi}
\label{Neraly}
\end{equation}

from which it follows from ~(\ref{EigenvalueEqn}) that

\begin{equation}
{(i\partial_{+}-
{g^{2}\over\sqrt{2}}\sigma)\psi=(\partial_{+}\phi)\psi.}
\label{Meraly}
\end{equation}

So, if $\psi$ satisfies (20) and $\phi$ ~(\ref{Meraly}), 
then $\tilde{\psi}=e^{i\phi}\psi$ will satisfy
(3). We can write

\begin{equation}
\psi=
{ e^{i\varphi(x^{-})}e^{i\phi} \over 
2\sqrt{2\lambda+\partial_{-}^{2}f^{3}}} \left[
\matrix{ 2\lambda+\partial_{-}^{2}f^{3} \cr
\partial_{-}^{2}f^{1}+i\partial_{-}^{2}f^{2}}\right]
\label{quark}
\end{equation}

where $\varphi(x^{-})$ is arbitrary and,

\begin{equation}
\partial_{+}\phi =-{g^{2}\over 2\sqrt{2}}\Big(
{f^{i}\partial^{2}_{-}f^{i}+2\lambda f^{3}\over 
2\lambda+\partial_{-}^{2}f^{3}}\Big).
\label{phase}
\end{equation}

It is not difficult to obtain an action in terms of $f_{i}$, 
$i=1,2,3$ and $\phi$ which yields equations of motion
equivalent to $(6)$ and $(24)$. It is not significantly
different from the action introduced in~\cite{AntPap} to treat
spin degrees of freedom. Then, the continuos symmetry referred
to above is not apparent, and one might consider fixing it at
the classical level. It is not clear to us that this should be
the correct approach, although this issue clearly deserves
further study beyond the scope of this communication.
We limit ourselves to presenting classical
solutions with properties which are physically acceptable. We do
this in the next section.

5) One way in which our soliton solution ~(\ref{CurveSoln}) can be
made to aquire the properties expected of a $SU(2)$ baryon, is
to consider the reparametrization

\begin{equation}
\partial_{-}s=sech^{2}(as)
\label{sprimed}
\end{equation}

i.e.

\begin{equation}
f^{i}=f^{i}(s(x^{-}));\quad \partial^{2}_{-}f^{i}=(\partial^{2}_{-}s)
f^{i}_{,s}+(\partial_{-}s)^{2}f^{i}_{,ss}
\label{indfs}
\end{equation}

Although the expressions ~(\ref{indfs}) are rather involved and will 
not be 
displayed in this communication, we note that

\begin{equation}
\int dx^{-}\psi^{\dagger}_{a}\psi_{a}=\int dx^{-}\lambda (x^{-})=\int 
{dx^{-}\over 2}\sqrt{\partial_{-}^{2}f^{i}\partial_{-}^{2}f^{i}} =
\int^{\infty}_{-\infty}{ds\over 2\partial_{-}s} 
\sqrt{(\partial_{-}s)^{4}\kappa^{2}(s)+(\partial_{-}^{2}s)^{2}}=2.
\label{BaryonNumber}
\end{equation}

Due to the improved behaviour at $x^{-}\to\pm\infty$, the $SU(N)$ 
color
charges vanish and

\begin{equation}
P^{+}=-{g^{2}\over 4}\int dx^{-}Tr(\partial_{-}^{2}\sigma)\sigma=
{g^{2}\over 4}\int dx^{-}Tr(\partial_{-}\sigma )^{2}={g^{2}\over 8}
\int dx^{-}(\partial_{-}s)^{2}={g^{2}\over 4a}
\label{energy}
\end{equation}

We have verified that 
$P^{-}=i\int dx^{-}\psi_{a}^{\dagger}\partial_{-}\psi_{a}=0.$ 
This is a non-trivial calculation, requiring detailed knowledge 
of the solution
~(\ref{quark})-~(\ref{indfs}). 

6) We now illustrate how 'tHooft's solution to the large $N$ gap 
equation
can be constructed from one of our classical $SU(2)$ solutions. A 
related
analysis has been carried out by Papanicoloau~\cite{Pap} for the 
classical
solutions of the $O(2)$ Gross-Neveu model~\cite{NevPap}. Papanicoloau
obtains a system of pseudospin constraints for fermionic systems with 
$O(N)$ symmetry and he describes how to construct equal time 
correlators 
which satisfy the large $N$ limit of these constraints as well as the 
theory's Heisenberg equations of motion from finite $N$ solutions to 
the
classical equations of motion. $QCD_{2}$ is of course $U(N_{C})$ 
symmetric
but more importantly, we will consider unequal time correlators
\footnote{Reference~\cite{wadia} discusses $U(N)$ symmetric theories 
in an
equal time formalism.}. In other words, we are able to construct the  
"masterfield" for unequal time correlators from solutions to the 
classical
equations of motion.

If one denotes the standard $U(N)$ invariant "time" ordered two point
by $G(x;y)$ and rescales as usual $\psi_{a}\to\sqrt{N}\psi_{a}$,
$g^{2}\to{g^{2}\over N}$, one obtains in the large $N$ limit the 
standard gap
equation

\begin{equation}
i\sqrt{2}\partial^{x}_{+}G(x;y)+g^{2}\int dw^{-}G(x^{+},x^{-
};x^{+},w^{-})
|x^{-}-w^{-}|G(x^{+},w^{-};y^{+},y^{-})=i\delta(x^{-}-y^{-}).
\label{gapeqn}
\end{equation}

Integrating both sides of ~(\ref{gapeqn}) with respect to $x^{+}$ 
infinitesimally around $y^{+}$, we obtain

\begin{equation}
i\sqrt{2}\quad
lim_{\epsilon\to 0}\Big[ G(x^{-},y^{+}+\epsilon;y^{-},y^{+})-
G(x^{-},y^{+}-\epsilon;y^{-},y^{+})]=i\delta (x^{-}-y^{-}).
\label{lnc}
\end{equation}

In a perturbative second quantized formalism, ~(\ref{lnc}) is an 
anticommutation relation. 

We consider now the solutions of constant curvature 
~(\ref{SecondCurveSoln})
with $\kappa=2k_{-}$. In the terminology of our previous section, 
'tHooft's
meson background corresponds to "static" ($x^{+}$ independent) 
solutions. It
is a peculiarity of $QCD_{2}$ in the chiral limit and in light cone 
quantization that the static solutions of the equations of motion 
~(\ref{EqnMot}) are also their $g^{2}\to 0$ limit. From equations
$(19)$ and $(23)$ it then follows that

\begin{equation}
\left[\matrix{\varphi_{1k_{-}}\cr\varphi_{2k_{-}}}\right]= 
\sqrt{{\kappa\over 2}}\left[\matrix{e^{ik_{-}x^{-}}e^{i\pi/2}
\cr e^{-ik_{-}x^{-}}e^{-i\pi/2}}\right].
\label{qwffcc}
\end{equation}

The relativistic bilocal field is constructed from a superposition of 
these
solutions as \footnote{The need for halving the number of color 
degrees of
freedom in order to satisfy Papanicoloau's pseudo spin constraints in 
an 
equal "time" treatment of $QCD_{2}$ has been anticipated by A.Jevicki
~\cite{PC}.}

\begin{eqnarray}
\nonumber
G(x,y)=\theta (x^{+}-y^{+})\int_{0}^{\infty}{dk_{-}\over 2\pi}
f_{+}(k_{-})e^{-i\omega_{k_{-}}(x^{+}-y^{+})}\varphi_{1k_{-}}(x^{-})
\varphi^{*}_{1k_{-}}(y^{-}) \\
-\theta (y^{+}-x^{+})\int_{0}^{\infty}{dk_{-}\over 2\pi}
f_{-}(k_{-})e^{i\omega_{k_{-}}(x^{+}-y^{+})}\varphi_{2k_{-}}(x^{-})
\varphi^{*}_{2k_{-}}(y^{-}).
\label{lna}
\end{eqnarray}

The functions $f_{\pm}$ are chosen so that ~(\ref{lnc}) is satisfied. 
A
simple choice is 

\begin{equation}
f_{+}(k_{-})=f_{-}(k_{-})={\sqrt{2}\over k_{-}}.
\label{lpos}
\end{equation}

It is now straight forward to substitute (32)-(33) in the gap equation
(29) which is satisfied provided

\begin{equation}
\omega_{k_{-}}={g^{2}\over 2}\int {dk_{-}'\over 2\pi}
{\epsilon (k_{-}+k_{-}')\over k_{-}'^{2}},
\label{too}
\end{equation}

which is the familiar (tachyonic) zero quark mass limit of 'tHooft's
background dispersion~\cite{tHooft}. We remark that if one was to
regard 'tHooft's large $N_{c}$ bilinears $G(x,y)$ as built form the
"time" dependant ansatz $\varphi (x^{-})e^{-i\omega_{k}x^{+}},$ then
these do not satisfy the classical equations of motion in agreement 
with
conclusions presented in [17].

\vspace{1cm}

{\it Acknowledgements} RdMK would like to 
thank the members of the Brown High Energy 
Physics group, where part of this work was 
done, for their hospitality.

%\vspace*{-0.5cm}

\end{document}